# When Developers Build Security Tools for Themselves: Requirement Misalignment in Automation Ecosystems

Oluwafolajinmi Adewole
*York University, Canada*
fj2005@yorku.ca

Elmira Onagh
*York University, Canada*
eonagh@yorku.ca

Maleknaz Nayebi
*York University, Canada*
mnayebi@yorku.ca

***Abstract*—Security automation tools are often built by developers for other developers, creating the expectation that requirements are well understood and correctly implemented, given their shared expertise. However, this assumption does not hold in practice. This paper studies how security and privacy requirements are expressed, implemented, and distributed in the GitHub Actions ecosystem. Analyzing 196 developer discussions and 998 reusable security tools, we find a consistent misalignment: some frequently discussed requirements are weakly implemented, while others are widely implemented yet remain unsatisfactory in practice. In addition, requirement support is unevenly distributed across domains, concentrating on detection-oriented capabilities while leaving concerns such as reliability, portability, and performance underrepresented. These findings suggest that requirement failure in developer-built ecosystems is not due to a lack of knowledge but to how requirements are implicitly encoded and fragmented across a two-sided platform. Requirements are not captured in a single specification, but distributed across workflows, tool artifacts, and developer discourse. We argue that requirements engineering must move beyond system-level specification toward ecosystem-level reasoning. In developer-built platforms, requirements are implicit, distributed, and continuously reinterpreted. The key challenge is not identifying requirements, but preventing known requirements from being lost, fragmented, or distorted as they are realized across reusable tools.**

***Index Terms*—requirements engineering, software ecosystems, security automation, requirement realization, requirement misalignment**

## I. Introduction

Security automation tools are increasingly used by developers to detect vulnerabilities, scan dependencies, identify leaked secrets, and enforce secure development practices. Many of these tools are built by developers for other developers, which creates a natural expectation: if the builders and users belong to the same technical community, then the resulting tools should reflect developer security and privacy requirements well. Yet, in practice, important requirements remain fragmented, weakly operationalized, or persistently unsatisfied [1]. This paper argues that this is not only a tooling problem, but an emerging Requirements Engineering (RE) challenge. In modern development ecosystems, requirements are no longer realized only within a single software system. They are increasingly realized through collections of reusable artifacts that developers discover, compose, and depend on. GitHub Actions provides a particularly compelling setting for studying this shift. It is not merely a repository of automation components, but a two-sided platform: one side consists of developers and organizations who create reusable Actions, while the other consists of developers and teams who adopt them to automate their workflows. Figure 1 demonstrates this multi-sided ecosystem. Value emerges through interaction across these two sides. Tool builders compete for visibility, reuse, and trust; tool adopters seek solutions that match their practical needs. In principle, such a platform should accelerate alignment between requirements and implementation. In practice, however, the platform can also amplify distortion, fragmentation, and redundancy in requirements.

This tension is especially visible in security automation. GitHub Actions hosts a large and growing ecosystem of reusable tools, including scanners, analyzers, policy checkers, and other security-oriented automation components. Because Actions are published, reused, and composed on a common platform, they form a distributed requirement realization environment rather than a single engineered solution. Some requirements are encoded repeatedly across many tools, some are implemented only partially, and others are largely left implicit until developers encounter failures in practice. As a result, the platform does not simply provide security support; it shapes which security requirements become visible, implementable, and reusable.

From an RE perspective, this raises a broader question: *what happens when requirements are operationalized through a two-sided automation ecosystem rather than specified and validated within one target system?* Traditional RE assumes that requirements are elicited from stakeholders, translated into system properties, and validated within a defined implementation boundary [2]. In developer-built tool ecosystems, this boundary breaks down. Developers frequently encounter a security issue, automate a solution, and publish it for reuse, expecting that similar needs exist across the ecosystem. In principle, this should lead to cumulative coverage of requirements. In practice, however, requirement realization remains uneven. Requirements are not explicitly captured but implicitly encoded across YAML workflows, marketplace artifacts, feature claims, and developer discussions. They are distributed across the platform rather than systematically specified, mak-ing it difficult to align expectations and implementation.

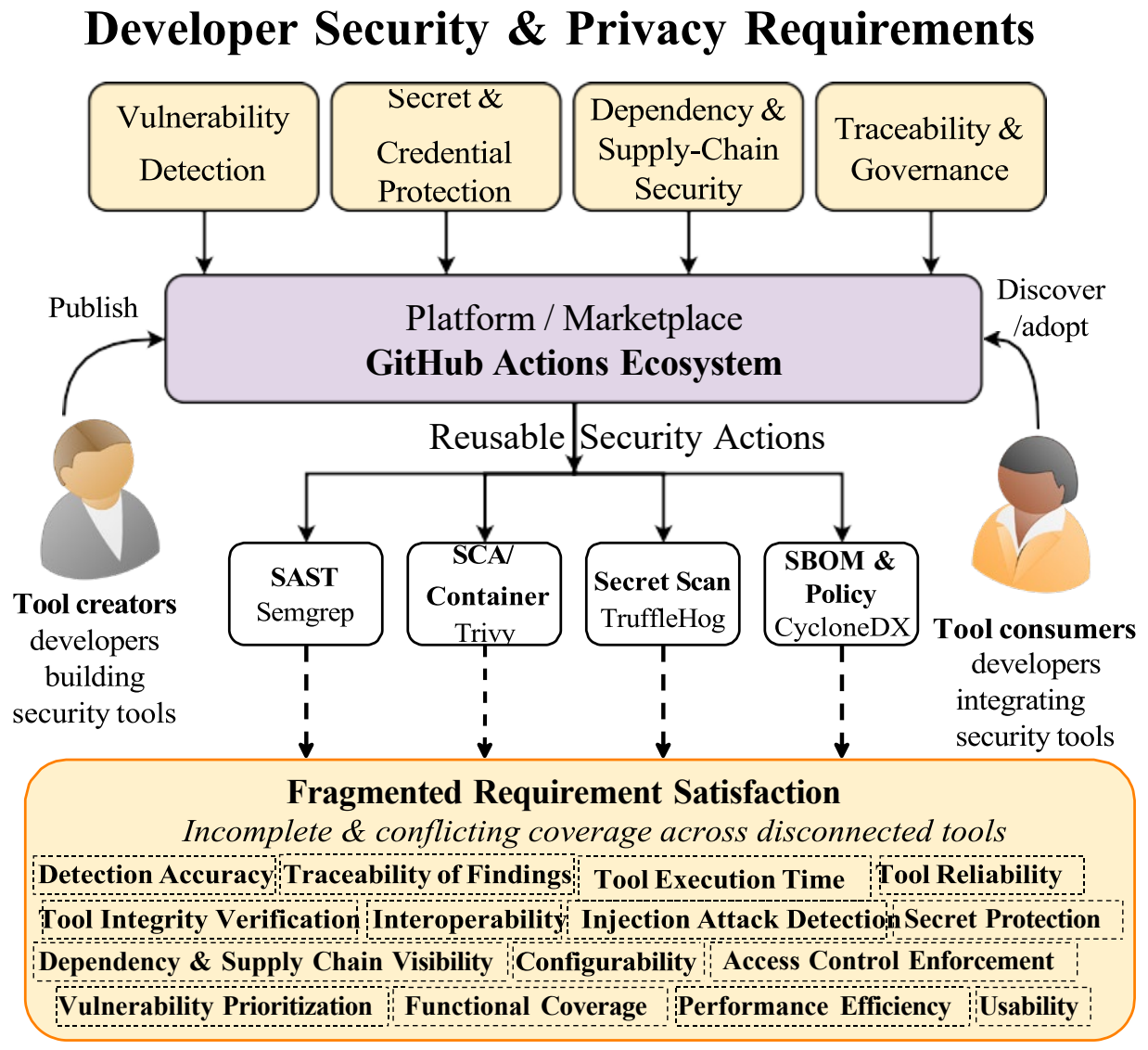

Fig. 1. The overview of the two-sided marketplace for software security on the GitHub marketplace

In this paper, we use the GitHub Actions security ecosystem as a motivating case for a broader RE!Next vision. Our empirical observations show that even when developers know the *problem space* well and build tools for one another, requirements gaps persist. Some requirements are widely discussed among developers but weakly operationalized in tools. Others are widely implemented yet continue to generate frustration, suggesting that presence does not imply satisfaction. We argue that these patterns reveal a new RE challenge: reasoning about the prominence, operationalization, and satisfaction of requirements in developer-built platform ecosystems. To explore this challenge, we ask:

**RQ1:** What security and privacy requirements are most needed in developers' discussions of security automation tools?

**RQ2:** How broadly are these requirements operationalized in reusable security automation tools?

**RQ3:** How is requirement operationalization distributed across security domains, and where does it become uneven or fragmented?

The central claim of this paper is that requirement failure in such ecosystems is not mainly due to a lack of expertise. Rather, it is rooted in the way requirements are distributed across a two-sided platform, where expectations, implementations, and claims evolve independently. This makes GitHub Actions more than a repository of automation logic. It is a platform in which requirements are negotiated indirectly through reuse, adoption, complaints, and tool evolution. For RE, this suggests the need to move beyond system-centered thinking toward ecosystem-level methods that can reason about requirement salience, operationalization, and satisfaction across reusable artifacts.

The remainder of this paper is structured as follows. Section II introduces GitHub Actions and related work. Section III presents an example linking requirements to tool behavior. Section IV describes our study design and data. Section V reports findings for RQ1–RQ3. We conclude our paper with implications for requirements engineering in ecosystem settings in Section VI.

## II. Background and Related Work

GitHub Actions, referred to as Actions, is a workflow automation framework integrated into GitHub. Developers define workflows in YAML files stored in the `.github/workflows` directory, and these workflows are triggered by events such as commits, pull requests, or scheduled jobs [3], [4]. A workflow is composed of jobs and steps, and steps can invoke reusable components called *Actions*. These components encapsulate automation logic such as testing, building, dependency analysis, or secret scanning, and can be created by repository owners or reused from the GitHub Marketplace [3], [5], [6]. We situate our contribution within this wider research context, drawing on its methodological perspectives, empirical investigations, and supporting research artifacts [7]–[78].

As introduced above, GitHub Marketplace functions as a two-sided platform of tool publishers and tool adopters [6]. Similar platform dynamics have been studied in mobile app stores, where release and marketing decisions shape which capabilities become visible to users [79], [80]. In principle, such a platform should help align tool capabilities with developer needs: when a developer encounters a recurring problem, they may automate the solution, publish it, and expect other developers facing the same problem to reuse it. Over time, one would expect the ecosystem to cover a substantial portion of common requirements.

However, prior work suggests this expectation does not hold in security automation. Although Actions are widely adopted [3], [4], [81], security-related Actions remain a relatively small part of the ecosystem [3], [4]. This is concerning because the security risks surrounding Actions are growing: 18% of custom GitHub Actions have vulnerable dependencies [82], 77.4% of JavaScript-based Actions depend on at least one vulnerable `npm` package release [83], and recent supply-chain compromises have affected thousands of repositories at once [84]–[87]. A growing body of empirical work characterizes how developers adopt and reuse Actions and how the ecosystem evolves over time [3], [4], yet it does not examine the security and privacy requirements developers report in practice (**RQ1**). A separate line of work examines security issues in Actions and proposes mitigations [5], [88], but it remains largely vulnerability and risk-driven: it focuses on identifying and preventing security flaws rather than tracing how developer requirements are implemented across reusable Actions (**RQ2**). More broadly, RE research on open-commercial ecosystems shows that requirements emerge and are clarified through distributed, continuous interactions, but rarely measures how developer-reported requirements are realized across reusable automation artifacts and how that realization is distributed across domains (**RQ3**) [89].

From a requirements engineering perspective, this is an ecosystem-level problem. Requirements are not realized by one system or one development team, but across many reusable artifacts contributed by different actors on the platform. A developer may expect a security tool to provide *Detection Accuracy*, *Dependency & Supply Chain Visibility*, *Vulnerability Prioritization*, and *Usability*, but these expectations may be distributed unevenly across multiple tools, each implementing only a fragment of the requirement space. This motivates our study: we examine (i) which requirements are most frequently raised by developers, (ii) how broadly those requirements are implemented across reusable security Actions, and (iii) how that implementation is distributed across security domains.

## III. Illustrative Example: Requirement Gaps in a Security Action

*Snyk* is a widely used security tool available through the GitHub Marketplace.[1] It performs dependency scanning and reports discovered vulnerabilities as part of an automated workflow [6]. This Action is included in our dataset and serves as an example of how security requirements are encoded in reusable automation artifacts.

In a typical workflow, Snyk is invoked on each code push to scan dependencies and report vulnerabilities within the development pipeline:

```yaml
name: Snyk Security Scan
on: [push]

jobs:
  security:
    runs-on: ubuntu-latest
    steps:
      - name: Checkout repository
        uses: actions/checkout@v3

      - name: Run Snyk to check for vulnerabilities
        uses: snyk/actions/node@master
        env:
          SNYK_TOKEN: ${{ secrets.SNYK_TOKEN }}
        with:
          command: test
```

This workflow encodes several security requirements directly in a reusable automation artifact. It specifies when the scan is triggered (on push), what is analyzed (project dependencies), and how the scan is configured and executed. While it operationalizes requirements such as vulnerability detection and supply chain visibility, it does not explicitly capture others, such as prioritization, contextualization, or reliability, that developers often expect in practice.

This example helps make our research questions concrete. A developer adopting such a tool is unlikely to think in terms of abstract categories. Instead, they bring practical expectations: the tool should reliably detect relevant vulnerabilities, cover important dependencies, prioritize findings so they can act on them, and integrate smoothly into their workflow. In our study, these expectations correspond closely to RQ1 categories such as *Detection Accuracy*, *Dependency & Supply Chain Visibility*, *Vulnerability Prioritization*, *Usability*, and *Traceability*.

At first glance, the Snyk-based workflow appears to satisfy these needs. It clearly implements some form of vulnerability scanning, which is exactly the kind of operationalization captured in RQ2. A reusable Action such as this encodes security requirements into executable workflow logic: it runs automatically, produces findings, and can be integrated into a development pipeline with only a few lines of YAML.

Yet the example also shows why implementation does not necessarily imply alignment. A team may expect *Detection Accuracy*, but still experience false positives, missing context, or incomplete results [90]. They may expect *Vulnerability Prioritization*, but still need to manually triage which findings matter [91]. They may expect *Usability* and *Traceability*, but find that integrating results into their issue management or review workflow requires extra manual effort [92], [93]. And they may expect broad *Dependency & Supply Chain Visibility*, but still experience dependencies and supply chain risks within their projects.

This makes the example useful for all three RQs. It illustrates the kinds of requirements developers most frequently discuss (RQ1), the fact that such requirements are indeed implemented in reusable tools (RQ2), and the central problem of the paper: requirement realization is often uneven, partial, or fragmented even when the tool appears to offer the right capability (RQ3). More broadly, the example shows why GitHub Actions is a compelling RE setting. The platform enables developers to package and share solutions quickly, but it does not guarantee that repeated publication leads to complete or well-aligned requirement coverage. In a two-sided automation platform, requirements are shaped not only by what tools claim to do, but also by how developers experience, interpret, and reuse those tools in practice.

## IV. Methodology

We use a mixed-method empirical design to examine the relationship between developer-expressed requirements. Consistent with the exploratory and visionary nature of an RE!Next paper, our goal is not to provide definitive ecosystem-wide claims, but to ground a broader requirements engineering perspective in empirical observations from a large and visible automation platform.

### A. Data Collection

We collected two complementary datasets: (i) a dataset of reusable security automation tools called Actions from GitHub Marketplace, and (ii) a dataset of developers' discussions from public online platforms.

**Security automation tools.** We selected GitHub Marketplace as the primary source of tool data because it is a two-sided platform [6]. This makes it an appropriate setting for studying how developer needs become encoded into reusable automation artifacts. We collected 998 GitHub Actions listed under the *security* category. For each Action, we retrieved its Marketplace description, repository metadata, and YAML

[1] https://github.com/marketplace/snyk

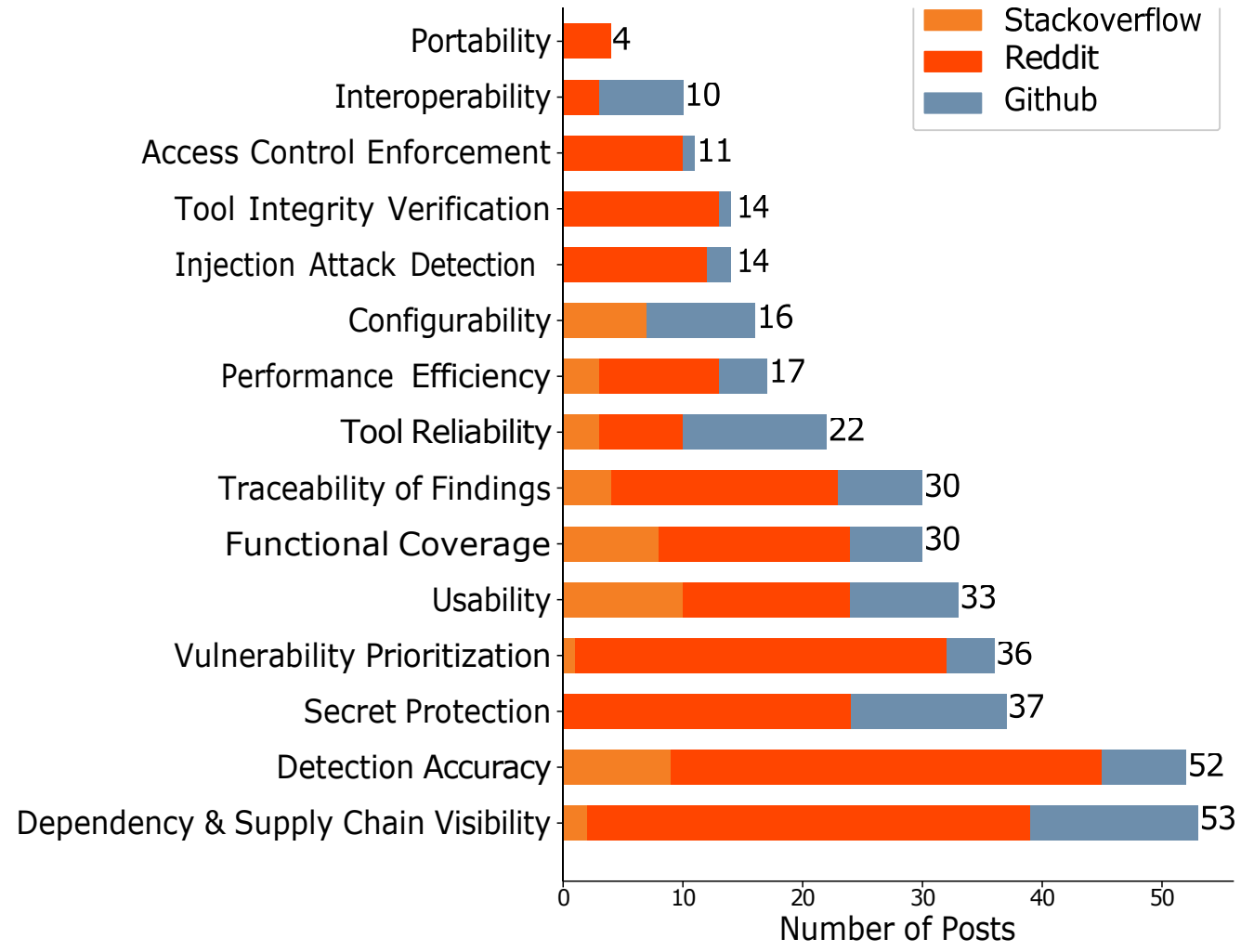

Fig. 2. Security requirements across developer discussions. Values give the number of posts per requirement (categories are non-mutually exclusive).

workflow-related artifacts.

**Developer discussions.** To capture developer expectations and frustrations, we collected public posts from Stack Overflow, Reddit, and GitHub using their official APIs [94]–[96]. Stack Overflow posts were retrieved using security-related tags such as static application security testing (`SAST`) and software composition analysis (`SCA`), Reddit posts were collected from security-relevant communities such as `r/sysadmin`, `r/programming`, and `r/netsec`, and GitHub data consisted of issues and related discussion artifacts from public security-tool repositories such as `trivy`, `gitleaks`, and `codeql-action`. We manually screened the retrieved candidates and retained only posts that substantively discuss security automation tools in a development workflow context, discarding off-topic or promotional posts. The final discussion dataset contained 35 Stack Overflow posts, 113 Reddit posts, and 48 GitHub entries.

## B. Exploratory Design

**RQ1.** To identify the security and privacy requirements most salient in developer discussions, we performed thematic coding [97] over the discussion dataset. The first author developed an initial codebook inspired by security concerns reflected in Open Worldwide Application Security Project (OWASP) guidance [98], then iteratively refined it through open coding on a subset of 50 posts. The resulting scheme was stabilized into 15 requirement categories. Each post was then read in full and annotated with one or more requirement categories by two researchers, with disagreements resolved through discussion until consensus was reached. We treat these posts as signals of the prominence of requirements because they capture developers' expressed concerns, expectations, complaints, and workarounds in real-world usage contexts rather than in formal specification documents.

**RQ2.** We then extracted security-relevant features (security-relevant capabilities stated in an Action's artifacts) from the 998 collected Actions, following the methodology of [28]. To do this, we fine-tuned a `Llama 3` model on a dataset of 100 manually labeled examples, each pairing a GitHub Action description with its capabilities mapped to the 15 requirement categories, with labeling performed by the first author. We then applied the fine-tuned model to all Actions, using their YAML-related descriptions and repository metadata as input, which yielded 837 unique features. To assess extraction quality, the first author manually reviewed a random sample of 50 Actions, verified that the extracted features were accurate and correctly mapped, and corrected errors before analysis. While we do not report inter-rater agreement, this level of manual validation was sufficient for the exploratory purpose of identifying ecosystem-level patterns in a visionary paper. Next, we assigned Actions to 13 security domains based on [99]. We then performed domain classification using a separately prompted `Llama 3` instance, which was provided with the domain definitions and each Action's description as input. Because an Action may support multiple forms of security functionality, domain assignment was multi-label rather than exclusive. The first author manually reviewed a sample of 50 classified Actions and corrected inconsistencies where necessary. We subsequently analyzed requirement coverage across domains by computing, for each requirement–domain pair, the proportion of Actions within the domain that operationalize the given requirement.

# V. Results

In this section, we summarize our findings across 196 security-relevant discussions and 837 security Actions to answer the research questions.

## A. *RQ1: What security and privacy requirements are most salient in developers' discussions of security automation tools?*

Following the coding procedure in Section IV, we analyzed 196 public posts from Stack Overflow, Reddit, and GitHub. Figure 2 shows that the most salient requirements are not primarily about the mere presence of security features, but about whether tools are trustworthy, actionable, and usable in real workflows. The two most frequently expressed concerns are *Dependency & Supply Chain Visibility* (27.0%) and *Detection Accuracy* (26.5%). Developers repeatedly questioned whether tools can detect malicious or vulnerable dependencies reliably, and whether findings can be trusted given false positives, false negatives, and weak contextualization. A second cluster of salient requirements concerns how findings are interpreted and acted on. *Secret Protection* (18.9%), *Vulnerability Prioritization* (18.4%), *Usability* (16.8%), *Functional Coverage* (15.3%), and *Traceability of Findings* (15.3%) all appear prominently in developer discussions. These results show that developers do not simply expect tools to detect problems; they expect them to support decision making, reduce noise, and fit coherently into broader development workflows.

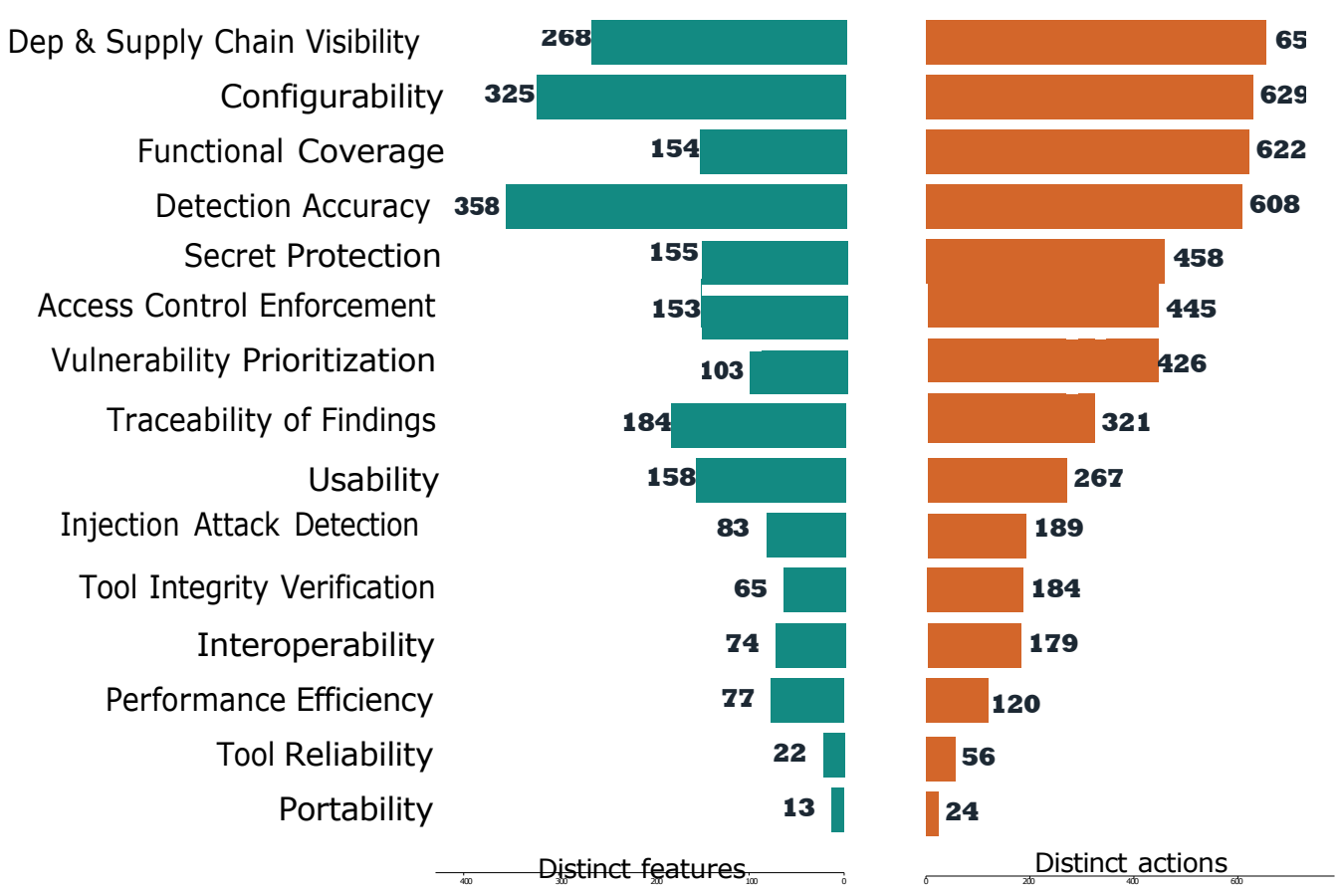

Fig. 3. Operationalization breadth per requirement: distinct features (left) and distinct Actions (right), sorted by Action count ascending.

TABLE I
DISTRIBUTION OF GITHUB ACTIONS ACROSS SECURITY DOMAINS

| Security Domain | # Actions | Percentage |
|---|---|---|
| Static Application Security Testing (SAST) | 395 | 39.6% |
| Secret Management | 366 | 36.7% |
| Software Composition Analysis (SCA) | 305 | 30.6% |
| Infrastructure as Code Security (IaC Security) | 198 | 19.8% |
| Interactive Application Security Testing (IAST) | 195 | 19.5% |
| Dynamic Application Security Testing (DAST) | 89 | 8.9% |
| Container Security Scanning | 74 | 7.4% |
| Threat Modeling | 33 | 3.3% |
| Security Information and Event Management (SIEM) | 12 | 1.2% |
| Web Application Firewall (WAF) | 12 | 1.2% |
| Intrusion Detection/Prevention Systems (IDS/IPS) | 6 | 0.6% |
| Hybrid Application Security Testing (HAST) | 4 | 0.4% |
| Runtime Application Self-Protection (RASP) | 4 | 0.4% |

A third group of requirements reflects operational and non-functional concerns. *Tool Reliability* (11.2%), *Performance Efficiency* (8.67%), *Configurability* (8.16%), *Injection Attack Detection* (7.14%), *Tool Integrity Verification* (7.14%), *Access Control Enforcement* (5.61%), *Interoperability* (5.10%), and *Portability* (2.04%) occur less frequently, but they reveal important expectations about robustness, integration, and trust in the tools themselves. Notably, these concerns are increasingly evident in emerging contexts such as AI-assisted development, where developers question whether tools can effectively detect vulnerabilities introduced by automation and whether the tools themselves can be considered trustworthy.

> Developer security requirements are not narrow or purely technical. They span detection, workflow fit, actionability, reliability, and trust. More importantly, they are expressed reactively: developers articulate them when tools fall short. For an RE perspective, this is significant because it shows that requirement salience in developer-built tool ecosystems is often revealed through friction in use rather than through explicit upfront specification.

### *B. RQ2: How broadly are these requirements operationalized in reusable security automation tools?*

To examine how developer-expected security requirements are operationalized in practice, we examined 998 Actions and 837 features described in Section IV. We measure breadth along two complementary dimensions: (i) the number of distinct features implementing a requirement, and (ii) the number of distinct Actions in which those features appear. Together, these dimensions distinguish requirements that are merely present from those that are implemented at scale and in diverse ways. Table I lists the 13 security domains of Actions [99]. Figure 3 shows that requirement operationalization is highly uneven across the ecosystem. A small subset of requirements is implemented broadly and repeatedly. *Dependency and Supply Chain Visibility* is the most broadly operationalized requirement, appearing in 655 Actions and expressed through 268 distinct features. *Configurability* (629 Actions, 325 features), *Functional Coverage* (622 Actions, 154 features), and *Detection Accuracy* (608 Actions, 358 features) also exhibit substantial breadth. These categories are not only common but also encoded through a wide range of capabilities, indicating strong ecosystem attention. At the opposite end, several requirements remain thinly operationalized. *Portability* is addressed by only 24 Actions and 13 distinct features, making it the narrowest requirement on both dimensions. *Tool Reliability* (56 Actions, 22 features) and *Performance Efficiency* (120 Actions, 77 features) are similarly weak in breadth. *Interoperability* and *Tool Integrity Verification* also remain limited relative to other broadly operationalized categories.

This unevenness is also reflected in the domain structure. As shown in Table I, the ecosystem is heavily concentrated in a small number of domains, such as SAST, Secret Management, and SCA. These domains account for much of the observed breadth. The results show that reusable security automation is primarily oriented toward *detection*, *scanning*, and *pre-deployment analysis*. In contrast, domains such as SIEM, WAF, IDS/IPS, HAST, and RASP remain sparsely represented.

To examine how evenly requirements are implemented across domains, we computed coverage as the percentage of Actions within each domain that support a given requirement. Figure 4 shows that a few requirements, including *Dependency & Supply Chain Visibility*, *Detection Accuracy*, *Configurability*, *Functional Coverage*, and *Access Control Enforcement*, achieve relatively high coverage across many domains. In contrast, requirements such as *Performance Efficiency*, *Tool Reliability*, and especially *Portability* remain weakly covered across most domains. Thus, breadth is not only selective at the ecosystem level but also unevenly distributed within domains.

> Reusable security automation tools do not operationalize developer requirements uniformly. Instead, the ecosystem broadly encodes a subset of highly visible, detection-oriented requirements while leaving others, particularly those related to robustness, integration, and operational continuity, comparatively under-realized.

### *C. RQ3: Where the security capabilities mismatch the stated requirements?*

RQ1 captures what developers state they need, expressed through recurring concerns, complaints, and workarounds in public discussions. RQ2 captures what the ecosystem operationalizes, expressed through extracted features, Action counts, and requirement coverage. We identify two distinct forms of mismatch when these two sides diverge. A *coverage gap* occurs when a requirement is highly salient in developer discussions but is only weakly operationalized across Actions. A *satisfaction gap* occurs when a requirement is broadly operationalized, yet developers continue to discuss it as a persistent problem, suggesting that numerical presence does not translate into perceived adequacy.

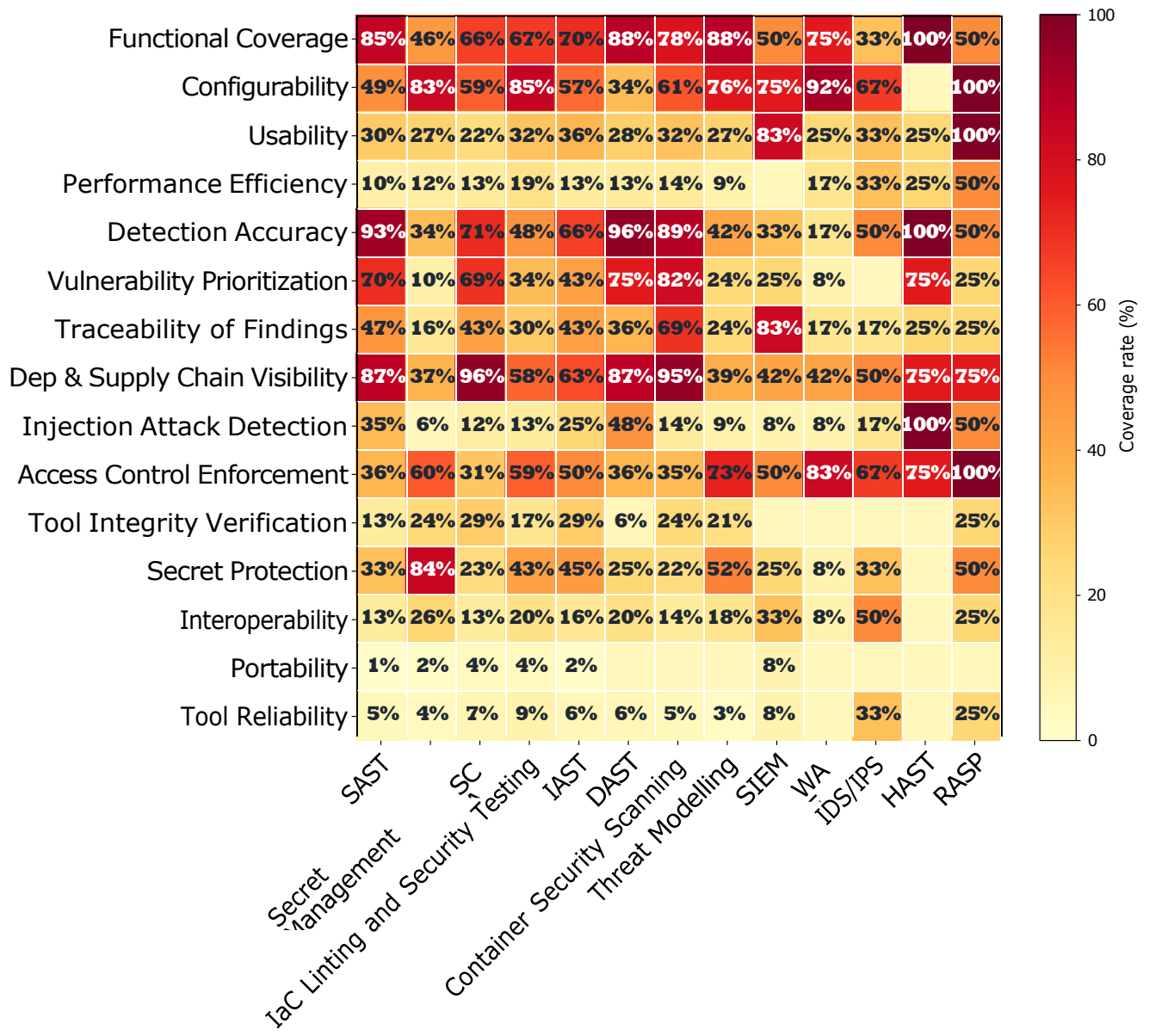

Fig. 4. Requirement coverage rate across security domains (% of each domain's Actions operationalizing each requirement).

*Coverage Gaps:* Several requirements appear as clear coverage gaps. *Performance Efficiency* is one of the strongest examples: developers repeatedly describe tools that slow down or block CI/CD workflows, yet only 120 of 998 Actions operationalize this requirement. *Tool Reliability* is similarly weak, with only 56 Actions and 22 distinct features, despite recurring developer frustration with tools that hang, fail silently, or behave inconsistently. *Portability* is the narrowest requirement in the ecosystem, addressed by only 24 Actions across six domains, even though developers explicitly raise concerns about vendor lock-in and the cost of switching tools. Other categories point to emerging areas of under-realization. *Injection Attack Detection* and *Tool Integrity Verification* are increasingly reflected in developer concerns, particularly in settings involving AI-assisted development and tool composition, yet remain only moderately or sparsely supported in the current ecosystem. This suggests that the operationalization of requirements is not keeping pace with the evolving nature of developer security expectations.

*Satisfaction Gaps:* Other requirements show the opposite pattern: they are broadly operationalized, yet continue to appear prominently in developer discussions. *Detection Accuracy* is the clearest case. It is addressed by 608 Actions and expressed through 358 distinct features, making it one of the most broadly operationalized requirements in the ecosystem. Yet it remains one of the most noticeable developer concerns, reflecting frustration with false positives, missed vulnerabilities, and low-confidence findings. The problem here is not absence, but inadequacy of requirements implementation. This asymmetry between implementing a requirement and satisfying it reinforces prior findings in release planning, where satisfaction and dissatisfaction are non-symmetric constructs that must be balanced explicitly [100]. A similar pattern appears for *Vulnerability Prioritization*. Although 426 Actions support this requirement, developers still struggle to identify which findings actually matter and often have to go through many alerts that turn out to be irrelevant or not worth acting on. The same issue appears for *Functional Coverage*, *Traceability*, *Usability*, and *Secret Protection* [101]. While many tools claim to support these requirements, developers continue to report gaps in coverage, unclear results, limited explanations, and incomplete protection.

Taken together, the ecosystem does not fail uniformly. Some requirements are genuinely under-supported, while others are numerically well represented but qualitatively unsatisfactory. This distinction matters. Adding more tools that claim *Detection Accuracy* will not resolve a satisfaction gap if those tools continue to produce noisy or context-poor results. Likewise, developers cannot be expected to adapt in the first place.

From an RE perspective, the key insight is that requirement realization in developer-built tool ecosystems is not only incomplete, but misaligned. Some requirements are missing in breadth, while others are present in volume but fail in satisfaction. This distinction is between reusable tool ecosystems.

> Even among technically knowledgeable developers, requirement realization across the ecosystem is neither uniform nor aligned. Some requirements remain under-supported, while others are widely implemented yet fail to meet developer expectations. Requirement presence does not imply requirement adequacy. Developer-built tool ecosystems may appear functionally rich, yet still fail to satisfy core requirements, showing a fundamental limitation of system-centered RE approaches in reusable, ecosystem-driven settings.

### D. Implications and RE!Next Vision

The findings from RQ1–RQ3 reveal a structural challenge for requirements engineering in developer-built tool ecosystems. Despite being technically sophisticated and directly experiencing the problems, developers are unable to consistently encode their own requirements into reusable tools. The gap between expectation and realization persists even when the builders and users are the same community.

*The Proximity Paradox:* A natural assumption is that developers building tools for other developers will encode relevant requirements effectively. Our results challenge this. Requirements such as *Detection Accuracy*, *Traceability*, and *Usability* are among the most frequently discussed concerns (RQ1) and also among the most broadly operationalized capabilities (RQ2), yet dissatisfaction remains high (RQ3). This indicates that shared expertise does not guarantee precise requirement articulation. We refer to this as the *proximity paradox*: when builders and users share domain knowledge, requirements are often left implicit. What is considered obvious is rarely specified with enough precision to guide implementation. As a result, tools implement a version of the requirement, but not the one developers expect in practice.

*Implicit Requirements as the Default:* RQ1 shows that requirements are rarely expressed through formal specifications, but instead emerge through complaints, workarounds, and post-hoc discussions. Developers typically adopt tools, encounter limitations during use, and only then articulate unmet needs. This makes requirements systematically implicit and reactive. Prior work observed the same pattern in mobile ecosystems, where requirements, including requests to remove functionality, surface through user discourse after release rather than upfront specification [102]. RQ2 and RQ3 further show that even when requirements are implemented at scale, their meaning remains ambiguous. A tool claiming *Detection Accuracy* may satisfy one interpretation, while developers expect another (e.g., fewer false positives, or better prioritization). Because these interpretations are not explicitly captured, mismatches persist.

*Implications for Requirements Engineering:* These findings suggest that traditional RE assumptions, such as explicit elicitation, stable specifications, and system-level validation, do not hold in developer-built tool ecosystems. Instead, requirements are distributed, implicit, and realized collectively across many independently developed artifacts. This motivates a shift toward ecosystem-level RE, where: (i) developer discourse (e.g., posts) is treated as a primary source of requirements, (ii) requirement evaluation distinguishes between presence and satisfaction, and (iii) Requirement operationalization is analyzed across tools rather than within a single system.

More broadly, this challenge extends beyond security tools to any ecosystem of reusable artifacts, including package repositories, plugin architectures, and AI toolchains. Requirements are not owned by a single system, but emerge from the interaction of many loosely coordinated components. This aligns with longitudinal evidence that practitioners favor good-enough solutions over optimal ones in release planning [51-54], suggesting that ecosystem actors are unlikely to converge on complete requirement coverage without explicit support. The central insight of this work is that requirement failure in these ecosystems is not due to a lack of knowledge but due to misalignment in how known requirements are articulated, distributed, and operationalized. Addressing this requires new RE methods that explicitly account for implicit requirements, evolving expectations, and ecosystem-level realization.

### E. Threats to Validity

This work is exploratory and has several limitations.

**Construct validity.** The requirement categories, feature extraction, and domain assignments rely on a combination of manual interpretation and model-assisted analysis, with validation performed on sampled data rather than through full multi-rater agreement.

**Internal validity.** Our mismatch analysis uses developer post frequency and Action-level feature counts as proxies for requirement salience and operationalization; these measures capture broad patterns, but not implementation quality or true user satisfaction.

**External validity.** The study is scoped to GitHub Actions and to public discussions from Stack Overflow, Reddit, and GitHub, and therefore does not cover other automation ecosystems or all developer communities. These limitations are consistent with the goal of this RE!Next paper: to ground an emerging requirements engineering challenge rather than to provide exhaustive or fully generalizable measurements.

## VI. Conclusion

We studied how security and privacy requirements are expressed, implemented, and distributed in a developer-built automation ecosystem. Our findings show a consistent misalignment: some requirements are highly salient but weakly operationalized, while others are widely implemented yet remain unsatisfactory. In addition, requirement support is unevenly distributed across domains, concentrating on detection while leaving other concerns under-realized. These results challenge a core assumption that developers building tools for themselves will naturally encode requirements well. Instead, even with shared expertise, requirement gaps persist.

From an RE perspective, this reflects a shift from system-level specification to ecosystem-level realization. Requirements are no longer captured in a single artifact, but distributed across workflows, tools, and developer discussions. As a result, alignment cannot be ensured through traditional elicitation and specification alone. We argue that RE must evolve to reason about requirements as implicit, distributed, and platform-mediated. *The key challenge is not identifying requirements, but ensuring that known requirements are consistently and adequately realized across reusable artifacts.* This challenge extends beyond security tools to any developer-built ecosystem. Understanding and addressing this misalignment is a central problem for the next generation of requirements engineering.

## References

[1] R. N. Rajapakse, M. Zahedi, and M. A. Babar, "An empirical analysis of practitioners' perspectives on security tool integration into devops," in *Proceedings of the 15th ACM / IEEE International Symposium on Empirical Software Engineering and Measurement (ESEM)*, ser. ESEM '21. New York, NY, USA: Association for Computing Machinery, 2021. [Online]. Available: https://doi.org/10.1145/3475716.3475776

[2] B. Nuseibeh and S. Easterbrook, "Requirements engineering: a roadmap," in *Proceedings of the Conference on The Future of Software Engineering*, ser. ICSE '00. New York, NY, USA: Association for Computing Machinery, 2000, p. 35–46. [Online]. Available: https://doi.org/10.1145/336512.336523

[3] A. Decan, T. Mens, P. Rostami Mazrae, and M. Golzadeh, "On the use of github actions in software development repositories," in *2022 IEEE International Conference on Software Maintenance and Evolution (ICSME)*. IEEE, 2022, pp. 235–245. [Online]. Available: https://doi.org/10.1109/ICSME55016.2022.00029

[4] T. Kinsman, M. Wessel, M. A. Gerosa, and C. Treude, "How do software developers use GitHub actions to automate their workflows?" in *Proceedings of the 2021 IEEE/ACM 18th International Conference on Mining Software Repositories (MSR)*, 2021, pp. 420–431. [Online]. Available: https://doi.org/10.1109/MSR52588.2021.00054

[5] H. O. Delicheh and T. Mens, "Mitigating security issues in github actions," in *Proceedings of the 2024 ACM/IEEE 4th International Workshop on Engineering and Cybersecurity of Critical Systems (EnCyCriS) and 2024 IEEE/ACM Second International Workshop on Software Vulnerability*, ser. EnCyCriS/SVM '24. New York, NY, USA: Association for Computing Machinery, 2024, p. 6–11. [Online]. Available: https://doi.org/10.1145/3643662.3643961

[6] S. G. Saroar, W. Ahmed, E. Onagh, and M. Nayebi, "Github marketplace for automation and innovation in software production," *Information and Software Technology*, vol. 175, p. 107522, 2024. [Online]. Available: https://www.sciencedirect.com/science/article/pii/S0950584924001277

[7] O. Adewole, E. Onagh, and M. Nayebi, "When developers build security tools for themselves: Requirement misalignment in automation ecosystems," in *34th IEEE International Requirements Engineering Conference (RE)*. IEEE, 2026, pp. 534–541. [Online]. Available: https://doi.org/10.1109/RE68928.2026.00056

[8] F. Ahmed, X. Tan, F. Adewole, S. Datta, and M. Nayebi, "Do multimodal llms understand programming screenshots? inferring questions and extracting relevant content," *Empirical Software Engineering*, vol. 31, no. 4, p. 88, 2026. [Online]. Available: https://doi.org/10.1007/s10664-026-10804-8

[9] E. Onagh, D. Johnson, M. Vautrin, A. Davoodi, and M. Nayebi, "Good enough over optimal: A longitudinal case study on software release planning in digital health," *Journal of Systems and Software*, vol. 240, p. 112939, 2026. [Online]. Available: https://www.sciencedirect.com/science/article/pii/S016412122600172X

[10] F. Pepe, V. Nardone, M. Nayebi, and M. D. Penta, "This is the replication package of the manuscript "an empirical study on the integration of scientific contributions into machine learning frameworks"," Zenodo, 2026, research dataset, version 1. [Online]. Available: https://doi.org/10.5281/zenodo.20443093

[11] ——, "This is the replication package of the manuscript "an empirical study on the integration of scientific contributions into machine learning frameworks"," Zenodo, 2026, research dataset, version 2. [Online]. Available: https://doi.org/10.5281/zenodo.21289904

[12] F. Ahmed, X. Tan, F. Adewole, S. Datta, and M. Nayebi, "Inferring questions from programming screenshots," in *22nd IEEE/ACM International Conference on Mining Software Repositories (MSR)*. IEEE, 2025, pp. 750–755. [Online]. Available: https://doi.org/10.1109/MSR66628.2025.00111

[13] U. A. Koana, Q. H. Le, S. Rahman, C. Carlson, F. Chew, and M. Nayebi, "Correction to: Examining ownership models in software teams," *Empirical Software Engineering*, vol. 30, no. 1, p. 4, 2025, publisher correction concerning the spelling of Shadikur Rahman; correction notice itself displays the earlier misspelling. [Online]. Available: https://doi.org/10.1007/s10664-024-10566-1

[14] E. Onagh and M. Nayebi, "Extension decisions in open source software ecosystem," *Journal of Systems and Software*, vol. 230, p. 112552, 2025. [Online]. Available: https://www.sciencedirect.com/science/article/pii/S0164121225002213

[15] E. Onagh, A. Davoodi, and M. Nayebi, "The impact of foundational models on patient-centric e-health systems," in *49th IEEE Annual Computers, Software, and Applications Conference (COMPSAC)*. IEEE, 2025, pp. 2014–2019. [Online]. Available: https://doi.org/10.1109/COMPSAC65507.2025.00281

[16] F. Pepe, C. Farkas, M. Nayebi, G. Antoniol, and M. D. Penta, "Replication package of the paper how do papers make into machine learning frameworks: A preliminary study on tensorflow (version 2)," Zenodo, 2025, research dataset, not a scholarly paper. [Online]. Available: https://doi.org/10.5281/zenodo.14203745

[17] ——, "How do papers make into machine learning frameworks: A preliminary study on tensorflow," in *33rd IEEE/ACM International Conference on Program Comprehension (ICPC)*. IEEE, 2025, pp. 1–6. [Online]. Available: https://doi.org/10.1109/ICPC66645.2025.00046

[18] J. Shin, C. Tang, T. Mohati, M. Nayebi, S. Wang, and H. Hemmati, "Prompt engineering or fine-tuning: An empirical assessment of llms for code," in *22nd IEEE/ACM International Conference on Mining Software Repositories (MSR)*. IEEE, 2025, pp. 490–502. [Online]. Available: https://doi.org/10.1109/MSR66628.2025.00082

[19] X. Tan, D. Yadav, F. Ahmed, and M. Nayebi, "Imager: Enhancing bug report clarity by screenshots," arXiv, 2025, arXiv version of the EASE 2025 paper; see Tan2025ImageREASE. [Online]. Available: https://doi.org/10.48550/arXiv.2505.01925

[20] X. Tan, D. Gengiti, and M. Nayebi, "Leveraging the power of images: Image recommendation to enhance issue reports," *Software: Practice and Experience*, vol. 55, no. 11, pp. 1785–1805, 2025. [Online]. Available: https://doi.org/10.1002/spe.70010

[21] S. P. Thirimanne, E. Y. Lemango, G. Antoniol, and M. Nayebi, "One documentation does not fit all: Case study of tensorflow documentation," in *49th IEEE Annual Computers, Software, and Applications Conference (COMPSAC)*. IEEE, 2025, pp. 1234–1243. [Online]. Available: https://doi.org/10.1109/COMPSAC65507.2025.00156

[22] X. Tan, D. Yadav, F. Ahmed, and M. Nayebi, "ImageR: Enhancing bug report clarity by screenshots," in *29th International Conference on Evaluation and Assessment in Software Engineering (EASE)*, 2025, conference publication confirmed by author; final proceedings DOI and page range not independently verified. [Online]. Available: https://conf.researchr.org/details/ease-2025/ease-2025-ai-models---data/18/ImageR-Enhancing-Bug-Report-Clarity-by-Screenshots

[23] F. Ahmed, S. Datta, and M. Nayebi, "Negative results of image processing for identifying duplicate questions on stack overflow," in *18th ACM/IEEE International Symposium on Empirical Software Engineering and Measurement (ESEM)*. ACM, 2024, pp. 281–291. [Online]. Available: https://doi.org/10.1145/3674805.3686688

[24] U. A. Koana, Q. H. Le, S. Rahman, C. Carlson, F. Chew, and M. Nayebi, "Examining ownership models in software teams: A systematic literature review and a replication study," *Empirical Software Engineering*, vol. 29, no. 6, p. 155, 2024. [Online]. Available: https://doi.org/10.1007/s10664-024-10538-5

[25] M. Nayebi, K. Kuznetsov, A. Zeller, and G. Ruhe, "Recommending and release planning of user-driven functionality deletion for mobile apps," *Requirements Engineering*, vol. 29, no. 4, pp. 459–480, 2024. [Online]. Available: https://doi.org/10.1007/s00766-024-00430-5

[26] M. Nayebi and B. Adams, "Image-based communication on social coding platforms," *Journal of Software: Evolution and Process*, vol. 36, no. 5, p. e2609, 2024. [Online]. Available: https://doi.org/10.1002/smr.2609

[27] M. Nayebi, H. Farrahi, and G. Ruhe, "Analysis of marketed versus not-marketed mobile app releases," arXiv, 2024, arXiv version of the 2016 RELENG conference paper; see Nayebi2016MarketedReleases. [Online]. Available: https://doi.org/10.48550/arXiv.2405.15752

[28] S. G. Saroar, W. Ahmed, E. Onagh, and M. Nayebi, "Github marketplace for automation and innovation in software production," *Information and Software Technology*, vol. 175, p. 107522, 2024. [Online]. Available: https://doi.org/10.1016/j.infsof.2024.107522

[29] U. A. Koana, F. Chew, C. Carlson, and M. Nayebi, "Ownership in the hands of accountability at brightsquid: A case study and a developer survey," in *31st ACM Joint European Software Engineering Conference and Symposium on the Foundations of Software Engineering (ESEC/FSE)*. ACM, 2023, pp. 2008–2019. [Online]. Available: https://doi.org/10.1145/3611643.3613890

[30] M. Nayebi, K. Kuznetsov, A. Zeller, and G. Ruhe, "User driven functionality deletion for mobile apps," in *31st IEEE International Requirements Engineering Conference (RE)*. IEEE, 2023, pp. 6–16. [Online]. Available: https://doi.org/10.1109/RE57278.2023.00011

[31] S. Rahman, F. Ahmed, and M. Nayebi, "Mining reddit data to elicit students' requirements during covid-19 pandemic," in *31st IEEE International Requirements Engineering Conference Workshops (REW)*. IEEE, 2023, pp. 76–84. [Online]. Available: https://doi.org/10.1109/REW57809.2023.00021

[32] S. G. Saroar and M. Nayebi, "Developers' perception of github actions: A survey analysis," in *27th International Conference on Evaluation and Assessment in Software Engineering (EASE)*. ACM, 2023, pp. 121–130. [Online]. Available: https://doi.org/10.1145/3593434.3593475

[33] M. Kuhrmann, P. Tell, R. Hebig, J. Klu¨nder, J. Mu¨nch, O. Linssen, D. Pfahl, M. Felderer, C. R. Prause, S. G. MacDonell, J. Nakatumba-Nabende, D. Raffo, S. Beecham, E. Tu¨zu¨n, G. Lo´pez, N. Paez, D. Fontdevila, S. A. Licorish, S. Ku¨pper, G. Ruhe, E. Knauss, O¨. O¨zcan-Top, P. M. Clarke, F. McCaffery, M. Genero, A. Vizca´ıno, M. Piattini, M. Kalinowski, T. Conte, R. Prikladnicki, S. Krusche,

A. Coskunçay, E. Scott, F. Calefato, S. Pimonova, R.-H. Pfeiffer, U. P. Schultz, R. Heldal, M. Fazal-Baqaie, C. Anslow, M. Nayebi, K. Schneider, S. Sauer, D. Winkler, S. Biffl, M. C. Bastarrica, and I. Richardson, "What makes agile software development agile?" *IEEE Transactions on Software Engineering*, vol. 48, no. 9, pp. 3523–3539, 2022. [Online]. Available: https://doi.org/10.1109/TSE.2021.3099532

[34] S. Rahman, U. A. Koana, and M. Nayebi, "Example driven code review explanation," in *ACM/IEEE International Symposium on Empirical Software Engineering and Measurement (ESEM)*. ACM, 2022, pp. 307–312. [Online]. Available: https://doi.org/10.1145/3544902.3546639

[35] S. G. Saroar, W. Ahmed, and M. Nayebi, "Github marketplace for practitioners and researchers to date: A systematic analysis of the knowledge mobilization gap in open source software automation," arXiv, 2022, early preprint with different title from later journal paper. [Online]. Available: https://doi.org/10.48550/arXiv.2208.00332

[36] M. Nayebi, G. Ruhe, and T. Zimmermann, "Mining treatment-outcome constructs from sequential software engineering data," *IEEE Transactions on Software Engineering*, vol. 47, no. 2, pp. 393–411, 2021. [Online]. Available: https://doi.org/10.1109/TSE.2019.2892956

[37] S. Wagner, D. M. Fernández, M. Felderer, A. Vetrò, M. Kalinowski, R. J. Wieringa, D. Pfahl, T. Conte, M.-T. Christiansson, D. Greer, C. Lassenius, T. Männistö, M. Nayebi, M. Oivo, B. Penzenstadler, R. Prikladnicki, G. Ruhe, A. Schekelmann, S. Sen, R. O. Spínola, A. Tuzcu, J. L. de la Vara, and D. Winkler, "Status quo in requirements engineering: A theory and a global family of surveys," in *Software Engineering 2021 (GI-Fachbereich Softwaretechnik)*, ser. Lecture Notes in Informatics, vol. P-310. Gesellschaft für Informatik e.V., 2021, pp. 115–116. [Online]. Available: https://doi.org/10.18420/SE2021_45

[38] Y. Hashemi, M. Nayebi, and G. Antoniol, "Documentation of machine learning software," in *27th IEEE International Conference on Software Analysis, Evolution and Reengineering (SANER)*. IEEE, 2020, pp. 666–667. [Online]. Available: https://doi.org/10.1109/SANER48275.2020.9054844

[39] M. Nayebi, "Eye of the mind: Image processing for social coding," in *42nd International Conference on Software Engineering, New Ideas and Emerging Results (ICSE-NIER)*. ACM, 2020, pp. 49–52. [Online]. Available: https://doi.org/10.1145/3377816.3381723

[40] A. Dekhtyar, J. H. Hayes, I. Hadar, E. Combs, A. Ferrari, S. Gregory, J. Horkoff, M. Levy, M. Nayebi, B. Paech, J. Payne, M. Primrose, P. Spoletini, S. Clarke, C. Brophy, D. Amyot, W. Maalej, G. Ruhe, J. Cleland-Huang, and D. Zowghi, "Requirements engineering (re) for social good: Re cares [requirements]," *IEEE Software*, vol. 36, no. 1, pp. 86–94, 2019. [Online]. Available: https://doi.org/10.1109/MS.2018.2874327

[41] W. Maalej, M. Nayebi, and G. Ruhe, "Data-driven requirements engineering: an update," in *41st International Conference on Software Engineering: Software Engineering in Practice (ICSE-SEIP)*. IEEE/ACM, 2019, pp. 289–290. [Online]. Available: https://doi.org/10.1109/ICSE-SEIP.2019.00041

[42] M. Nayebi, Y. Cai, R. Kazman, G. Ruhe, Q. Feng, C. Carlson, and F. Chew, "A longitudinal study of identifying and paying down architecture debt," in *41st International Conference on Software Engineering: Software Engineering in Practice (ICSE-SEIP)*. IEEE/ACM, 2019, pp. 171–180. [Online]. Available: https://doi.org/10.1109/ICSE-SEIP.2019.00026

[43] M. Nayebi and G. Ruhe, "Asymmetric release planning: Compromising satisfaction against dissatisfaction," *IEEE Transactions on Software Engineering*, vol. 45, no. 9, pp. 839–857, 2019. [Online]. Available: https://doi.org/10.1109/TSE.2018.2810895

[44] M. Nayebi, L. Dicke, R. Ittyipe, C. Carlson, and G. Ruhe, "Essmart way to manage customer requests," *Empirical Software Engineering*, vol. 24, no. 6, pp. 3755–3789, 2019. [Online]. Available: https://doi.org/10.1007/s10664-019-09721-w

[45] J. Perrie, J. Xie, M. Nayebi, M. Fokaefs, K. A. Lyons, and E. Stroulia, "City on the river: visualizing temporal collaboration," in *29th Annual International Conference on Computer Science and Software Engineering (CASCON)*. ACM, 2019, pp. 82–91. [Online]. Available: https://doi.org/10.5555/3370272.3370281

[46] F. Sarro and M. Nayebi, Eds., *Proceedings of the 3rd ACM SIGSOFT International Workshop on App Market Analytics, WAMA@ESEC/FSE 2019*. ACM, 2019, edited workshop proceedings, not a coauthored paper. [Online]. Available: https://doi.org/10.1145/3340496

[47] S. Wagner, D. M. Fernández, M. Felderer, A. Vetrò, M. Kalinowski, R. J. Wieringa, D. Pfahl, T. Conte, M.-T. Christiansson, D. Greer, C. Lassenius, T. Männistö, M. Nayebi, M. Oivo, B. Penzenstadler, R. Prikladnicki, G. Ruhe, A. Schekelmann, S. Sen, R. O. Spínola, A. Tuzcu, J. L. de la Vara, and D. Winkler, "Status quo in requirements engineering: A theory and a global family of surveys," *ACM Transactions on Software Engineering and Methodology*, vol. 28, no. 2, pp. 9:1–9:48, 2019. [Online]. Available: https://doi.org/10.1145/3306607

[48] J. H. Hayes, M. Nayebi, A. Dekhtyar, and B. Paech, "Re cares'18: First re cares workshop and event - re cares about giving back to alberta," in *26th IEEE International Requirements Engineering Conference (RE)*. IEEE Computer Society, 2018, pp. 436–438. [Online]. Available: https://doi.org/10.1109/RE.2018.00057

[49] M. Nayebi, H. Cho, and G. Ruhe, "App store mining is not enough for app improvement," *Empirical Software Engineering*, vol. 23, no. 5, pp. 2764–2794, 2018. [Online]. Available: https://doi.org/10.1007/s10664-018-9601-1

[50] M. Nayebi, "Data driven requirements engineering: Implications for the community," in *26th IEEE International Requirements Engineering Conference (RE)*. IEEE Computer Society, 2018, pp. 439–441. [Online]. Available: https://doi.org/10.1109/RE.2018.00058

[51] M. Nayebi, K. Kuznetsov, P. Chen, A. Zeller, and G. Ruhe, "Anatomy of functionality deletion: An exploratory study on mobile apps," in *15th International Conference on Mining Software Repositories (MSR)*. ACM, 2018, pp. 243–253. [Online]. Available: https://doi.org/10.1145/3196398.3196410

[52] M. Nayebi, S. J. Kabeer, G. Ruhe, C. Carlson, and F. Chew, "Hybrid labels are the new measure!" *IEEE Software*, vol. 35, no. 1, pp. 54–57, 2018. [Online]. Available: https://doi.org/10.1109/MS.2017.4541048

[53] M. Nayebi, "Analytical release management for mobile apps," Ph.D. dissertation, University of Calgary, 2018, doctoral dissertation; year confirmed by author.

[54] I. Hargreaves, D. Roth, M. R. Karim, M. Nayebi, and G. Ruhe, "Effective customer relationship management at ATB financial: A case study on industry-academia collaboration in data analytics," in *Highlighting the Importance of Big Data Management and Analysis for Various Applications*, ser. Studies in Big Data, M. Moshirpour, B. H. Far, and R. Alhajj, Eds. Springer, 2018, vol. 27, pp. 45–59, published online in 2017. [Online]. Available: https://doi.org/10.1007/978-3-319-60255-4_4

[55] S. M. D. A. Alam, M. Nayebi, D. Pfahl, and G. Ruhe, "A two-staged survey on release readiness," in *21st International Conference on Evaluation and Assessment in Software Engineering (EASE)*. ACM, 2017, pp. 374–383. [Online]. Available: https://doi.org/10.1145/3084226.3084254

[56] A. Hemmati, C. Carlson, M. Nayebi, G. Ruhe, and C. Saunders, "Analysis of software service usage in healthcare communication services," in *IEEE International Conference on Software Quality, Reliability and Security Companion (QRS-C)*. IEEE, 2017, pp. 565–566. [Online]. Available: https://doi.org/10.1109/QRS-C.2017.95

[57] S. J. Kabeer, M. Nayebi, G. Ruhe, C. Carlson, and F. Chew, "Predicting the vector impact of change - an industrial case study at brightsquid," in *ACM/IEEE International Symposium on Empirical Software Engineering and Measurement (ESEM)*. IEEE Computer Society, 2017, pp. 131–140. [Online]. Available: https://doi.org/10.1109/ESEM.2017.20

[58] M. Nayebi, H. Cho, H. Farrahi, and G. Ruhe, "App store mining is not enough," in *39th International Conference on Software Engineering Companion (ICSE)*. IEEE Computer Society, 2017, pp. 152–154. [Online]. Available: https://doi.org/10.1109/ICSE-C.2017.77

[59] M. Nayebi, H. Farrahi, and G. Ruhe, "Which version should be released to app store?" in *ACM/IEEE International Symposium on Empirical Software Engineering and Measurement (ESEM)*. IEEE Computer Society, 2017, pp. 324–333. [Online]. Available: https://doi.org/10.1109/ESEM.2017.46

[60] M. Nayebi and G. Ruhe, "Optimized functionality for super mobile apps," in *25th IEEE International Requirements Engineering Conference (RE)*. IEEE Computer Society, 2017, pp. 388–393. [Online]. Available: https://doi.org/10.1109/RE.2017.72

[61] M. Nayebi, M. Marbouti, R. Quapp, F. Maurer, and G. Ruhe, "Crowdsourced exploration of mobile app features: A case study of the fort mcmurray wildfire," in *39th IEEE/ACM International Conference on Software Engineering: Software Engineering in Society*

*(ICSE-SEIS)*. IEEE Computer Society, 2017, pp. 57–66. [Online]. Available: https://doi.org/10.1109/ICSE-SEIS.2017.8

[62] G. Ruhe, M. Nayebi, and C. Ebert, "The vision: Requirements engineering in society," in *25th IEEE International Requirements Engineering Conference (RE)*. IEEE Computer Society, 2017, pp. 478–479. [Online]. Available: https://doi.org/10.1109/RE.2017.70

[63] W. Maalej, M. Nayebi, T. Johann, and G. Ruhe, "Toward data-driven requirements engineering," *IEEE Software*, vol. 33, no. 1, pp. 48–54, 2016. [Online]. Available: https://doi.org/10.1109/MS.2015.153

[64] M. Nayebi, H. Farrahi, A. Lee, H. Cho, and G. Ruhe, "More insight from being more focused: analysis of clustered market apps," in *International Workshop on App Market Analytics (WAMA@FSE)*. ACM, 2016, pp. 30–36. [Online]. Available: https://doi.org/10.1145/2993259.2993266

[65] M. Nayebi, B. Adams, and G. Ruhe, "Release practices for mobile apps - what do users and developers think?" in *23rd IEEE International Conference on Software Analysis, Evolution, and Reengineering (SANER)*. IEEE Computer Society, 2016, pp. 552–562. [Online]. Available: https://doi.org/10.1109/SANER.2016.116

[66] G. Ruhe and M. Nayebi, "What counts is decisions, not numbers - toward an analytics design sheet," in *Perspectives on Data Science for Software Engineering*. Academic Press, 2016, pp. 111–114. [Online]. Available: https://doi.org/10.1016/B978-0-12-804206-9.00022-2

[67] M. Nayebi, H. Farrahi, and G. Ruhe, "Analysis of marketed versus not-marketed mobile app releases," in *Proceedings of the 4th International Workshop on Release Engineering (RELENG)*. ACM, 2016, pp. 1–4. [Online]. Available: https://doi.org/10.1145/2993274.2993281

[68] M. Nayebi, G. Ruhe, R. C. Mota, and M. Mufti, "Analytics for software project management - where are we and where do we go?" in *30th IEEE/ACM International Conference on Automated Software Engineering Workshops (ASEW)*. IEEE Computer Society, 2015, pp. 18–21. [Online]. Available: https://doi.org/10.1109/ASEW.2015.28

[69] M. Nayebi, K. Jantzen, and G. Ruhe, "Managing technical debt," Software Engineering Decision Support Lab, University of Calgary, Tech. Rep. SEDS 115/2015, 2015, confirmed by author; direct public file unavailable.

[70] M. Nayebi and K. Wnuk, "Summary of the 1st international workshop on open innovation in software engineering (oise 2015)," in *International Conference on Software and System Process (ICSSP)*. ACM, 2015, pp. 183–184. [Online]. Available: https://doi.org/10.1145/2785592.2785624

[71] M. Nayebi and G. Ruhe, "Trade-off service portfolio planning - a case study on mining the android app market," *PeerJ Preprints*, vol. 3, p. e1354, 2015, preprint. [Online]. Available: https://doi.org/10.7287/peerj.preprints.1354v1

[72] ——, "Analytical product release planning," in *The Art and Science of Analyzing Software Data*. Morgan Kaufmann/Elsevier, 2015, pp. 555–589. [Online]. Available: https://doi.org/10.1016/B978-0-12-411519-4.00019-7

[73] M. Nayebi, G. Ruhe, and B. Adams, "Tracking the evolution of android mobile apps by release cycle patterns - an empirical study," Software Engineering Decision Support Lab, University of Calgary, Tech. Rep. SEDS 114/2015, 2015, confirmed by author; direct public file unavailable.

[74] M. Nayebi, "The gandhi-washington method for extracting patterns from versioning histories," Software Engineering Decision Support Lab, University of Calgary, Tech. Rep. SEDS 113/2015, 2015, title and year confirmed by author; direct public file unavailable.

[75] M. Nayebi and G. Ruhe, "Analytical open innovation for value-optimized service portfolio planning," in *5th International Conference on Software Business (ICSOB)*, ser. Lecture Notes in Business Information Processing, vol. 182. Springer, 2014, pp. 273–288. [Online]. Available: https://doi.org/10.1007/978-3-319-08738-2_19

[76] M. Nayebi, "Feasibility study on android app store mining," Software Engineering Decision Support Lab, University of Calgary, Tech. Rep. SEDS 109/2013, 2014, confirmed by author; direct public file unavailable.

[77] M. Nayebi and G. Ruhe, "An open innovation approach in support of product release decisions," in *7th International Workshop on Cooperative and Human Aspects of Software Engineering (CHASE)*. ACM, 2014, pp. 64–71. [Online]. Available: https://doi.org/10.1145/2593702.2593709

[78] E. Onagh and M. Nayebi, "How specification explicitness shapes failure in llm-generated code," *arXiv preprint arXiv:2611.20195*, 2026.

[79] M. Nayebi, H. Farrahi, and G. Ruhe, "Analysis of marketed versus not-marketed mobile app releases," in *Proceedings of the 4th International Workshop on Release Engineering*, ser. RELENG 2016. New York, NY, USA: Association for Computing Machinery, 2016, p. 1–4. [Online]. Available: https://doi.org/10.1145/2993274.2993281

[80] M. Nayebi, H. Farahi, and G. Ruhe, "Which version should be released to app store?" in *2017 ACM/IEEE International Symposium on Empirical Software Engineering and Measurement (ESEM)*, 2017, pp. 324–333. [Online]. Available: https://doi.org/10.1109/ESEM.2017.46

[81] R. McCarthy and S. Berkovich, "How to harden github actions: An updated guide," https://www.wiz.io/blog/github-actions-security-guide, 2026, accessed: 2026-03-23.

[82] N. Dotan, "State of GitHub actions security report," https://info.legitsecurity.com/the-state-of-github-actions-security, 2023, accessed: 2026-03-23.

[83] H. Onsori Delicheh, A. Decan, and T. Mens, "Quantifying security issues in reusable javascript actions in github workflows," in *Proceedings of the 21st International Conference on Mining Software Repositories*, ser. MSR '24. New York, NY, USA: Association for Computing Machinery, 2024, p. 692–703. [Online]. Available: https://doi.org/10.1145/3643991.3644899

[84] Cybersecurity and Infrastructure Security Agency (CISA), "Supply chain compromise of third-party tj-actions/changed-files and reviewdog/action-setup@v1," https://www.cisa.gov/news-events/cybersecurity-advisories, 2025, accessed: 2026-03-23.

[85] B. Pariseau, "GitHub actions supply chain attack spotlights CI/CD risks," https://www.techtarget.com/searchitoperations/news/366621078/GitHub-Actions-supply-chain-attack-spotlights-CI-CD-risks, 2025, accessed: 2026-03-23.

[86] D. Rein, "Promptpwnd: Prompt injection vulnerabilities in github actions using ai agents," https://www.aikido.dev/blog/promptpwnd-github-actions-ai-agents, 2025, accessed: 2026-03-23.

[87] A. Ilgayev, "How we discovered vulnerabilities in ci/cd pipelines of popular open-source projects," https://cycode.com/blog/github-actions-vulnerabilities/, 2024, accessed: 2026-03-23.

[88] S. G. Saroar and M. Nayebi, "Developers' perception of github actions: A survey analysis," in *Proceedings of the 27th International Conference on Evaluation and Assessment in Software Engineering*, ser. EASE '23. New York, NY, USA: Association for Computing Machinery, 2023, p. 121–130. [Online]. Available: https://doi.org/10.1145/3593434.3593475

[89] E. Knauss, A. Yussuf, K. Blincoe, D. Damian, and A. Knauss, "Continuous clarification and emergent requirements flows in open-commercial software ecosystems," *Requirements Engineering*, vol. 23, 03 2018. [Online]. Available: https://doi.org/10.1007/s00766-016-0259-1

[90] O. Javed and S. Toor, "An evaluation of container security vulnerability detection tools," in *Proceedings of the 2021 5th International Conference on Cloud and Big Data Computing*, ser. ICCBDC '21. New York, NY, USA: Association for Computing Machinery, 2021, p. 95–101. [Online]. Available: https://doi.org/10.1145/3481646.3481661

[91] V. Koscinski, M. Nelson, A. Okutan, R. Falso, and M. Mirakhorli, "Conflicting scores, confusing signals: An empirical study of vulnerability scoring systems," in *Proceedings of the 2025 ACM SIGSAC Conference on Computer and Communications Security*, ser. CCS '25. New York, NY, USA: Association for Computing Machinery, 2025, p. 1904–1918. [Online]. Available: https://doi.org/10.1145/3719027.3765210

[92] Z. D. Wadhams, C. Izurieta, and A. M. Reinhold, "Barriers to using static application security testing (sast) tools: A literature review," in *Proceedings of the 39th IEEE/ACM International Conference on Automated Software Engineering Workshops*, ser. ASEW '24. New York, NY, USA: Association for Computing Machinery, 2024, p. 161–166. [Online]. Available: https://doi.org/10.1145/3691621.3694947

[93] O. Keskin, N. Gannon, B. Lopez, and U. Tatar, "Scoring cyber vulnerabilities based on their impact on organizational goals," in *2021 Systems and Information Engineering Design Symposium (SIEDS)*, 2021, pp. 1–6. [Online]. Available: https://doi.org/10.1109/SIEDS52267.2021.9483741

[94] Stack Exchange Inc., "Stack exchange api v2.3 documentation," https://api.stackexchange.com/docs, accessed: 2026-03-23.

[95] Reddit Inc., "Reddit api documentation," https://www.reddit.com/dev/api/, accessed: 2026-03-23.

[96] GitHub Inc., "Github rest api documentation," https://docs.github.com/en/rest, accessed: 2026-03-23.

[97] S. K. Ahmed, R. A. Mohammed, A. J. Nashwan, R. H. Ibrahim, A. Q. Abdalla, B. M. M. Ameen, and R. M. Khdhir, "Using thematic analysis in qualitative research," *Journal of Medicine, Surgery, and Public Health*, vol. 6, p. 100198, 2025. [Online]. Available: https://www.sciencedirect.com/science/article/pii/S2949916X25000222

[98] A. van der Stock, B. Glas, N. Smithline, T. Janca, and T. Gigler, "OWASP top 10: The Ten Most Critical Web Application Security Risks," https://owasp.org/Top10/2025/0x00_2025-Introduction/, 2025, accessed: 2026-03-23.

[99] J. Martelleur and A. Hamza, "Security tools in devsecops – a systematic literature review," Master's thesis, Linnaeus University, Sweden, 2023, diva2:1727554. [Online]. Available: https://lnu.diva-portal.org/smash/get/diva2:1727554/FULLTEXT01.pdf

[100] M. Nayebi and G. Ruhe, "Asymmetric release planning: Compromising satisfaction against dissatisfaction," *IEEE Transactions on Software Engineering*, vol. 45, no. 9, pp. 839–857, 2019. [Online]. Available: https://doi.org/10.1109/TSE.2018.2810895

[101] Y. Zhang, Y. Wu, T. Chen, T. Wang, H. Liu, and H. Wang, "How do developers talk about github actions? evidence from online software development community," in *Proceedings of the IEEE/ACM 46th International Conference on Software Engineering*. New York, NY, USA: Association for Computing Machinery, 2024. [Online]. Available: https://doi.org/10.1145/3597503.3623327

[102] M. Nayebi, K. Kuznetsov, A. Zeller, and G. Ruhe, "User driven functionality deletion for mobile apps," in *2023 IEEE 31st International Requirements Engineering Conference (RE)*, 2023, pp. 6–16. [Online]. Available: https://doi.org/10.1109/RE57278.2023.00011